\documentclass[aps,prb,twocolumn,showpacs,preprintnumbers,pra,superscriptaddress]{revtex4-1}
\usepackage{graphicx}  
\usepackage{dcolumn}   
\usepackage{bm}        
\usepackage{amssymb}   
\usepackage{framed}
\usepackage{amsmath}
\usepackage{hhline}
\usepackage{adjustbox}
\usepackage{subfigure,amsmath,verbatim,moreverb}
\usepackage{tabularx}
\usepackage{lipsum}
\usepackage{longtable}
\usepackage{booktabs}
\usepackage{rotating}
\usepackage[normalem]{ulem}
\usepackage{amsfonts}
\usepackage{booktabs}
\usepackage{color,soul}
\usepackage{etoolbox}
\AtBeginEnvironment{align}{\setcounter{subeqn}{0}}
\newcounter{subeqn} %

\begin{document}

\title{Correct and accurate polymorphic energy ordering of transition-metal monoxides obtained from semilocal and onsite-hybrid exchange-correlation approximations}
\author{Arghya Ghosh}
\email{ph17resch11006@iith.ac.in}
\affiliation{Department of Physics, Indian Institute of Technology, Hyderabad, India}
\author{Subrata Jana}
\email{Corresponding author: jana.18@osu.edu, subrata.niser@gmail.com}
\affiliation{Department of Chemistry \& Biochemistry, The Ohio State University, Columbus, OH 43210, USA}
\author{Manish K Niranjan}
\email{manish@phy.iith.ac.in}
\affiliation{Department of Physics, Indian Institute of Technology, Hyderabad, India}
\author{Fabien Tran}
\affiliation{Institute of Materials Chemistry, Vienna University of Technology, Getreidemarkt 9/165-TC, A-1060 Vienna, Austria}
\author{David Wimberger}
\affiliation{Institute of Materials Chemistry, Vienna University of Technology, Getreidemarkt 9/165-TC, A-1060 Vienna, Austria}
\author{Peter Blaha}
\affiliation{Institute of Materials Chemistry, Vienna University of Technology, Getreidemarkt 9/165-TC, A-1060 Vienna, Austria}
\author{Lucian A. Constantin}
\affiliation{Istituto di Nanoscienze, Consiglio Nazionale delle Ricerche CNR-NANO, 41125 Modena, Italy}
\author{Prasanjit Samal}
\affiliation{School of Physical Sciences, National Institute of Science Education and Research, HBNI, Bhubaneswar 752050, India}

\date{\today}

\begin{abstract}

The relative energetic stability of the structural phases of common antiferromagnetic transition-metal oxides (MnO, FeO, CoO, and NiO) within the semilocal and hybrid density functionals are fraught with difficulties. In particular, MnO is known to be the most difficult case for almost all common semilocal and hybrid density approximations. Here, we show that the meta-generalized gradient approximation (meta-GGA) constructed from the cuspless hydrogen model and Pauli kinetic energy density (MGGAC) can lead to the correct ground state of MnO. The relative energy differences of zinc-blende ($zb$) and rock-salt ($rs$) structures as computed using MGGAC are found to be in nice agreement with those obtained from high-level correlation methods like the random phase approximation or quantum Monte Carlo techniques. Besides, we have also applied the onsite hybrid functionals (closely related to DFT+$U$) based on GGA and meta-GGA functionals, and it is shown that a relatively high amount of Hartree-Fock exchange is necessary to obtain the correct ground-state structure. Our present investigation suggests that the semilocal MGGAC and onsite hybrids, both being computationally cheap, as methods of choice for the calculation of the relative stability of antiferromagnetic transition-metal oxides having potential applications in solid-state physics and structural chemistry.

\end{abstract}

\maketitle

\section{\label{introduction}Introduction}

Following the pioneering works of Hohenberg, Kohn, and Sham~\cite{hohenberg1964inhomogeneous,kohn1965self}, density functional theory (DFT) has become a highly successful and indispensable tool for studying the geometry and electronic structure of solid-state and condensed matter systems~\cite{burke2012perspective,RevModPhys.87.897,doi:10.1063/1.4869598}. Over the years, DFT-based first-principles studies have greatly contributed to the progress in various fields relevant to technology like nanoscience and molecular electronics. However, the accuracy and reliability of DFT depend crucially on the various approximations for the exchange-correlation (xc) energy functional~\cite{PerdewAIP01}, which includes all the many-body effects beyond the Hartree approximation. During the last few decades, several accurate approximations for the semilocal xc functionals, which are the computationally cheapest methods in DFT, have been proposed. However, their application to transition-metal oxides (TMOs) compounds having open $d$-shell still remains challenging~\cite{svane1990transition,peng2013polymorphic,schron2010energetic,schiller2015phase,peng2017synergy}

Here in this paper, we revisit the relative accuracy of different levels of xc methods to predict the ground-state properties of a prototypical open $d$-shell TMO, namely MnO, which is potentially very interesting in industrial applications, e.g., photoelectrochemical water splitting~\cite{kanan2012band,toroker2013transition}, solar energy conversion~\cite{peng2012semiconducting}, or magneto-piezoelectric effect~\cite{gopal2004first}. A large number of theoretical studies have been carried out for MnO phases~\cite{peng2013polymorphic,schron2010energetic,schiller2015phase,exchange2011archer,franchini2005density,peng2017synergy} that have led to the identification of different polymorphic phases~\cite{schron2010energetic}: rock-salt ($rs$), zinc-blende ($zb$), and wurtzite ($wz$). Concerning magnetism, MnO is antiferromagnetic (AF) with ferromagnetic planes stacked along the [111] (AF2) and [001] (AF1) directions for the $rs$ and $zb$ structures, respectively. Among the phases $rs$-AF2 and $zb$-AF1, the first one is found as the most stable one according to experiment \cite{RothPR58} as well as the random phase approximation (RPA) \cite{peng2013polymorphic} and diffuse Monte Carlo (DMC) methods \cite{schiller2015phase}, which are high-level \textit{ab initio} methods. However, this is not the case with the common generalized gradient approximations (GGA) and hybrid DFT xc methods that incorrectly predict the energy of the $zb$-AF1 phase to be lower than that of $rs$-AF2 \cite{peng2013polymorphic,schron2010energetic,schiller2015phase,peng2017synergy}. However, GGA+$U$ (GGA with a Hubbard $U$ correction) predicts the correct ground state of MnO, although a large and unphysical $U$ value is required \cite{kanan2012band}.

One may note that in addition to the aforementioned works, a very large number of other studies have investigated the ground-state or electronic properties of the $rs$-AF2 phase of MnO. A certain number of DFT and beyond DFT methods are used, and this includes many semilocal DFT methods (see Refs.~\onlinecite{peng2017synergy,zhang2020symmetry,sai2018evaluation,PhysRevMaterials.2.023802,tran2020shortcomings} for recent works), DFT+$U$ \cite{AnisimovPRB91,PhysRevB.86.115134}, various types of hybrids \cite{TranPRB06,MarsmanJPCM08,Liu_2019,jana2018efficient,jana2019screened,jana2020screened,jana2020improved}, the self-interaction corrected local density approximation \cite{svane1990transition}, the optimized effective potential method \cite{solovyev1998effective,EngelPRL09}, model Hamiltonian approach~\cite{exchange2011archer,erten2011theory,hossain2021hybrid,hossain2020transferability}, the quasi-particle $GW$ method \cite{massidda1997quasiparticle,Kotani_2008,rodl2009quasiparticle}, and dynamical mean field theory (DMFT) \cite{KunesNM08,MandalNCM19,mukherjee2014testing}. Nevertheless, it is important to mention that the most accurate of all these methods, namely, RPA, DMC, $GW$, and DMFT are computationally much more expensive than DFT-based methods.

Hence, from the point of view of efficiency, the preferred methods are the semilocal xc functionals in DFT. In particular, we mention that SCAN+rVV10+$U$, which consists of the strongly constrained and appropriately normed (SCAN) meta-GGA \cite{sun2015strongly} combined with the rVV10 van der Waals (vdW) \cite{SabatiniPRB13} functional and a Hubbard $U$ correction, has been found to perform well in case of TMOs~\cite{peng2017synergy}. It is clearly admitted that the meta-GGA functionals generally perform better than the GGAs in describing solid-state properties~\cite{sun2015strongly,jana2018assessment,patra2020way,jana2021accurate,jana2020accurate,jana2020insights,ghosh2021improved}. Concerning MnO, it has been shown that the polymorphic structural energy difference computed using SCAN+rVV10+$U$, where $U$ is determined from linear response theory, agrees very well with that obtained from DMC values~\cite{peng2017synergy}. Besides the SCAN-based methods, several other meta-GGA functionals are also proposed and tested for solid-state properties with consistently improved accuracy~\cite{tran2016rungs,sengupta2018from,shani2018accurate,peng2016versatile,mo2017assessment,jana2018assessment,jana2018assessing,jana2019improving,patra2019relevance,patra2019performance,patra2021correct,jana2021improved,ghosh2021improved}. Within those recent meta-GGAs, there are the functionals constructed from the cuspless hydrogen model [(r)MGGAC]~\cite{patra2019relevance,jana2021improved} that show potential promising accuracy for different challenging problems in solid-state physics~\cite{patra2021correct,jana2021improved,ghosh2021improved}. It is also quite an efficient semilocal functional that can predict band gaps of bulk and layered solids with reasonable accuracy~\cite{patra2020electronic, patra2021efficient,jana2021improved,tran2021bandgap,ghosh2021improved}.

Inspired by the promising performance of the (r)MGGAC functionals for solids, in the present paper we investigate the polymorphic energy ordering of MnO, FeO, CoO, and NiO using these methods, with a particular focus on MnO. Our results will also be compared with those obtained from other DFT methods and higher-level quantum methods.

This paper is organized as follows. In Sec.~\ref{methods} we first briefly describe the methods that we used for the calculations. Then, in Sec.~\ref{results} we present and discuss the results obtained for the polymorphic energy ordering of different TMOs, and Sec.~\ref{conclusions} presents the conclusions.

\section{\label{methods}Methods of calculation}

Belonging to the semilocal levels of approximation, the functionals PBE~\cite{perdew1996generalized}, SCAN~\cite{sun2015strongly}, r$^{2}$SCAN~\cite{furness2020accurate}, MGGAC~\cite{patra2019relevance}, and rMGGAC~\cite{jana2021improved} are considered for our calculations. The meta-GGA functionals (r$^{2}$)SCAN and (r)MGGAC, which depend on the density ($n=\sum_{i}|\psi_{i}|^2$), gradient of the density ($\nabla n$), and KS kinetic-energy density (KED) ($\tau=\frac{1}{2}\sum_{i}|\nabla\psi_{i}|^2$), are implemented in the generalized KS (gKS) \cite{SeidlPRB96,yang2016more,perdew2017understanding} scheme. The semilocal functionals can be expressed as
\begin{eqnarray}
 E_{\text{xc}} & = &\int\epsilon_{\text{xc}}(n,\nabla n,\tau)d^3r \nonumber\\
 &  = &\int\epsilon_{\text{x}}^{\text{LDA}}(n)F_{\text{xc}}(r_{s},s,\alpha^{\text{iso}})d^3r,
\label{Exc}
\end{eqnarray}
where $\epsilon_{\text{x}}^{\text{LDA}}=-\left(3/4\right)\left(3/\pi\right)^{1/3}n^{4/3}$ is the exchange energy density of the local density approximation (LDA), $F_{\text{xc}}$ is the xc enhancement factor (with no dependency on $\tau$ for GGAs), and $r_{s}=\left(3/\left(4\pi n\right)\right)^{1/3}$ is the Wigner-Seitz radius. Usually, the $\nabla n$- and $\tau$-dependencies of $F_{\text{xc}}$ are expressed via dimensionless variables, like the reduced density gradient $s=|\nabla n|/\left(2\left(3\pi^{2}\right)^{1/3}n^{4/3}\right)$ and the iso-orbital indicator $\alpha^{\text{iso}}=(\tau-\tau^{\text{W}})/\tau^{\text{UEG}}$, where $\tau^{\text{W}}$ is the von Weizs\"{a}cker KED and $\tau^{\text{UEG}}$ is the KED of the uniform electron gas (UEG). Actually, it is important to note that $\alpha^{\text{iso}}$ recognizes regions with single bonds, overlapping orbitals, and uniform density~\cite{PhysRevLett.111.106401,della2016kinetic}. 
\begin{figure}
\includegraphics[scale=0.63]{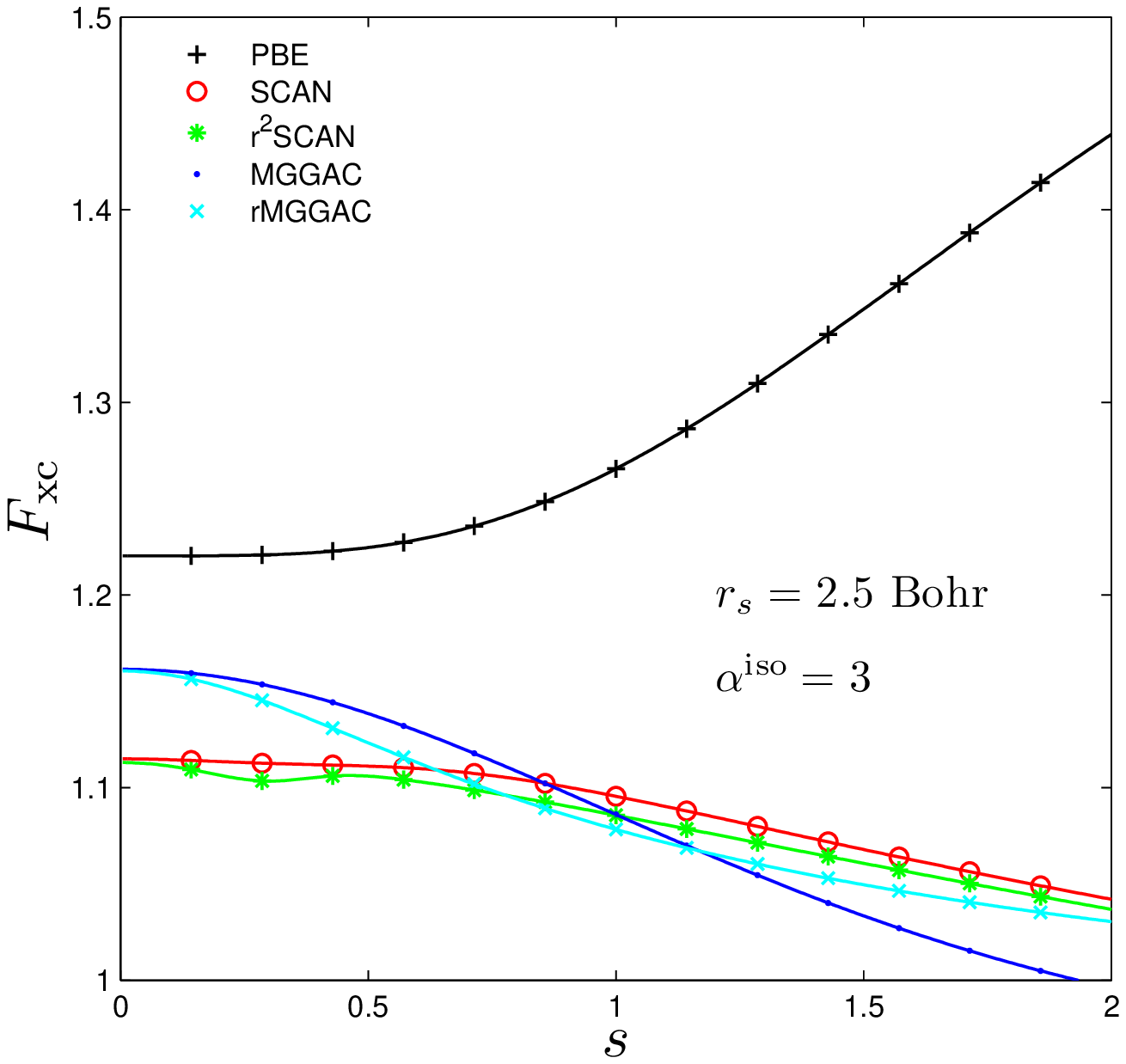}
\includegraphics[scale=0.63]{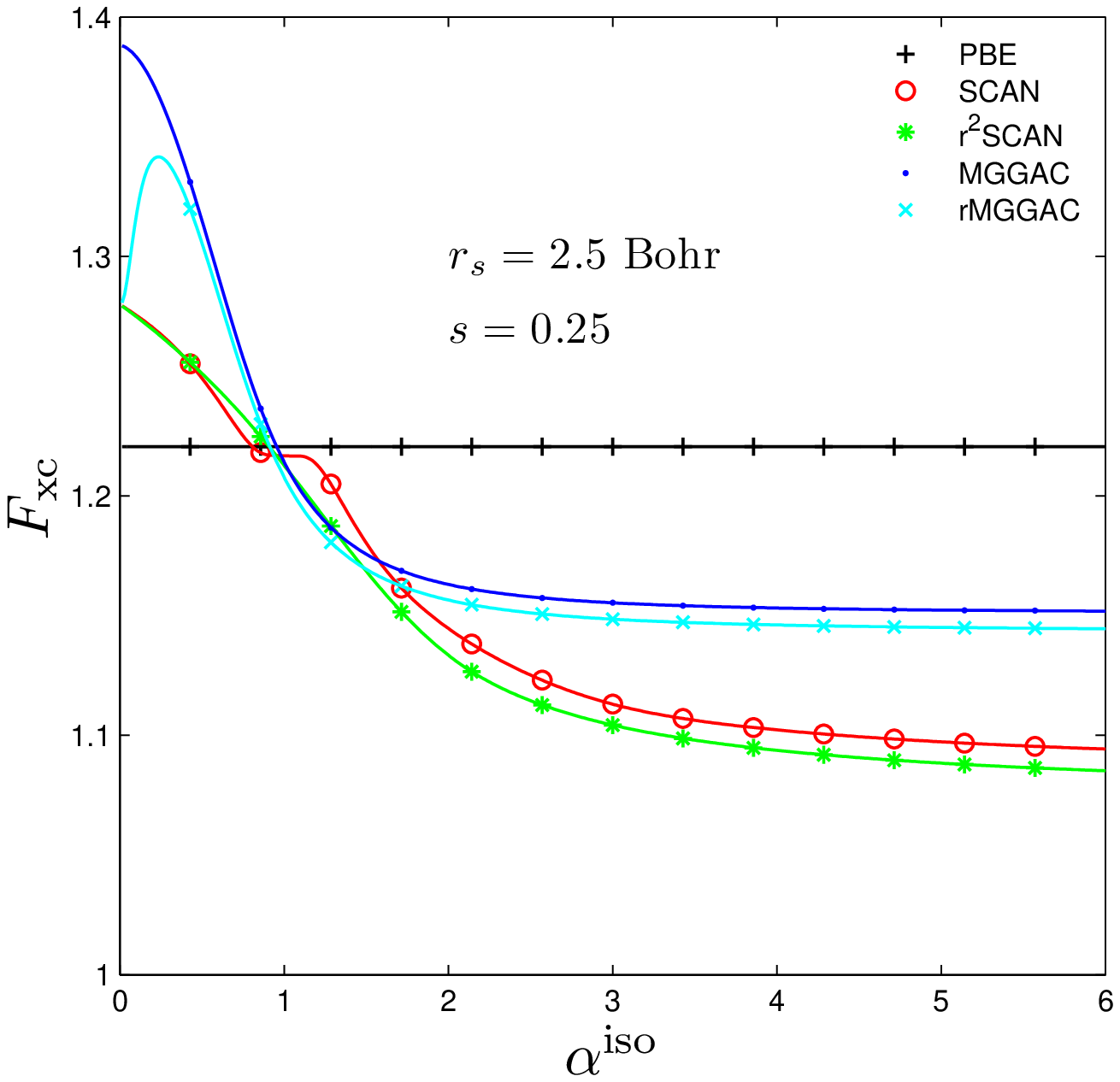}
\caption{\label{fig_Fxc}Enhancement factors $F_{\text{xc}}$ plotted as a function of $s$ (upper panel) and $\alpha^{\text{iso}}$ (lower panel). The value of the two other variables (that are kept fixed) are indicated in the respective panels. Note the different scales on the vertical axis.}
\end{figure}

Note that the SCAN functional was constructed such that it satisfies 17 exact mathematical constraints that can be satisfied by a meta-GGA. For instance, the exchange component of the xc enhancement factor, $F_{x}^{\text{SCAN}}$, recovers the exact fourth-order gradient approximation of exchange. r$^{2}$SCAN is a slightly modified version of SCAN that alleviates numerical problems encountered with SCAN. The development of the MGGAC functional is rather unusual and unconventional. The exchange part is based on the Becke-Roussel approach\cite{PhysRevA.39.3761}, and with a cuspless hydrogen exchange hole density~\cite{patra2019efficient}. MGGAC differs from SCAN for the following two reasons. First, only the exchange component of MGGAC is a meta-GGA, while the correlation part is a GGA. Second, $F_{x}^{\text{MGGAC}}$ depends only on $\alpha^{\text{iso}}$ (and not on $s$). However, the correlation part of rMGGAC is of the meta-GGA type \cite{jana2021improved}. It is worth to mention that (r$^{2}$)SCAN and (r)MGGAC respect the strongly tightened bound exchange ($F_x \leq 1.174$~\cite{perdew2014gedanken,sun2015strongly}) and possess ultranonlocality effects, which is important for the band gap problem~\cite{aschebrock2019ultranonlocality,patra2020electronic,patra2019relevance}. It may also be noted that MGGAC correlation is not free from the one-electron self-interaction error, whereas rMGGAC is. 

In Fig.~\ref{fig_Fxc}, the enhancement factor $F_{\text{xc}}$ is shown for the PBE, (r$^{2}$)SCAN, and (r)MGGAC functionals. They are plotted as functions of $s$ or $\alpha^{\text{iso}}$, with fixed chosen values for the other parameters. An obvious difference between the GGA PBE and the meta-GGAs concerns the sign of $\partial F_{\text{xc}}/\partial s$. While $\partial F_{\text{xc}}/\partial s$ is positive for PBE, it is negative for all four meta-GGAs. Note that in general the slope $\partial F_{\text{xc}}/\partial s$ of meta-GGAs can be positive or negative depending on the particular meta-GGA, but also, to a lesser extent, on the chosen fixed values of $r_{s}$ or $\alpha^{\text{iso}}$ (see Ref.~\onlinecite{tran2020shortcomings} for plots of $F_{\text{xc}}$ for other meta-GGAs). The other difference between PBE and the meta-GGAs concerns of course the variation with respect to $\alpha^{\text{iso}}$. Since PBE is a GGA, $\partial F_{\text{xc}}/\partial\alpha^{\text{iso}}=0$. The meta-GGAs have a negative value of $\partial F_{\text{xc}}/\partial\alpha^{\text{iso}}$ as for most meta-GGAs \cite{tran2020shortcomings}. As discussed in Ref.~\onlinecite{aschebrock2019ultranonlocality}, a more negative slope $\partial F_{\text{xc}}/\partial\alpha^{\text{iso}}$ leads to a larger derivative discontinuity, and consequently also to a larger band gap.

In addition to the aforementioned functionals, the onsite hybrid functionals, which are proposed specifically for systems with strongly correlated electrons~\cite{TranPRB06}, are also considered in this work. They share close similarities with the DFT+$U$ method, in particular since they are also applied only to the strongly correlated electrons, while the rest of the electrons are treated at the semilocal level. However, there are a few fundamental and technical differences. For instance, whereas $U$, which represents the screened Coulomb interaction, is the parameter on which the DFT+$U$ results depend, in the case of the onsite hybrids it is the amount of Hartree-Fock exchange that can be tuned to vary the results. Also, in the onsite hybrids the double counting correction is calculated at the semilocal DFT level, while a more ad hoc correction is used in DFT+$U$. The two onsite hybrid functionals that are tested in the present work, PBE-$\alpha$ and SCAN-$\alpha$, use PBE and SCAN as underlying semilocal functionals, respectively. For both, three different values of the amount of Hartree-Fock exchange $\alpha$ (0.25, 0.40, and 0.55) are used. Technical details about the onsite hybrids and their implementations can be found in Refs.~\onlinecite{NovakPSSB06,TranPRB06}. Note that in the SCAN-$\alpha$ calculations, the SCAN functional was applied non-self-consistently and the PBE potential was used instead.

\section{\label{results}Results}

\begin{table*}
\caption{Total energy difference $\Delta E=E_{zb}^{\text{AF1}}-E_{rs}^{\text{AF2}}$ (in meV/f.u.) between the $zb$-AF1 and $rs$-AF2 phases of MnO. A positive energy difference indicates that $rs$-AF2 is energetically more stable than $zb$-AF1 (as experimentally determined). The MGGAC, rMGGAC, r$^2$SCAN, and onsite hybrids results are calculated for this work. The results obtained with the other methods are from Refs.~\onlinecite{schiller2015phase,peng2013polymorphic,peng2017synergy}. $U$ and $V$ values are in eV.}
\begin{ruledtabular}
\begin{tabular}{llccccccccccc}
Functional type & Functional & $\Delta E$ \\ 
\hline
Semilocal&PBE&-244$^a$\\
&SCAN&-79$^a$\\
&r$^2$SCAN&-33\\
&MGGAC&118\\
&rMGGAC&101\\[0.2 cm]
Semilocal+vdW&PBE+TS	&	-151$^a$\\[0.2 cm]
Semilocal(+vdW)+$U$&PBE+$U$ ($U=3$)&-19$^{b}$\\
&PBE+TS+$U$ ($U=2$)	&	88$^{b}$\\
&SCAN+rVV10+$U$ ($U=2.8$)	&	138$^{a}$\\[0.2 cm]
Hybrid&HSE06&-28$^a$\\[0.2 cm]
Onsite hybrids&PBE-$\alpha$ ($\alpha=0.25$) & -84 \\
&PBE-$\alpha$ ($\alpha=0.40$) & -22   \\
&PBE-$\alpha$ ($\alpha=0.55$) & 23   \\
&SCAN-$\alpha$ ($\alpha=0.25$) & 38   \\
&SCAN-$\alpha$ ($\alpha=0.40$) & 98   \\
&SCAN-$\alpha$ ($\alpha=0.55$) & 145  \\[0.2 cm]
Higher level methods&PBE+$V{_{GW}}$ ($U=7$, $V=3$)	&	64$^{b}$\\
&RPA@PBE+$U$ ($U=3$)	&	67$^{b}$\\
&RPA@PBE+$V{_{GW}}$ ($U=7$, $V=3$)	&	131$^{b}$\\
&DMC	&	132$^{c}$
\end{tabular}
\end{ruledtabular}
\footnotetext[1]{Ref.~\onlinecite{peng2017synergy}.}
\footnotetext[2]{Ref.~\onlinecite{peng2013polymorphic}.}
\footnotetext[3]{Ref.~\onlinecite{schiller2015phase}.}
\label{tab-rel1}
\end{table*}

\begin{figure*}
\includegraphics[width=18 cm, height=14 cm]{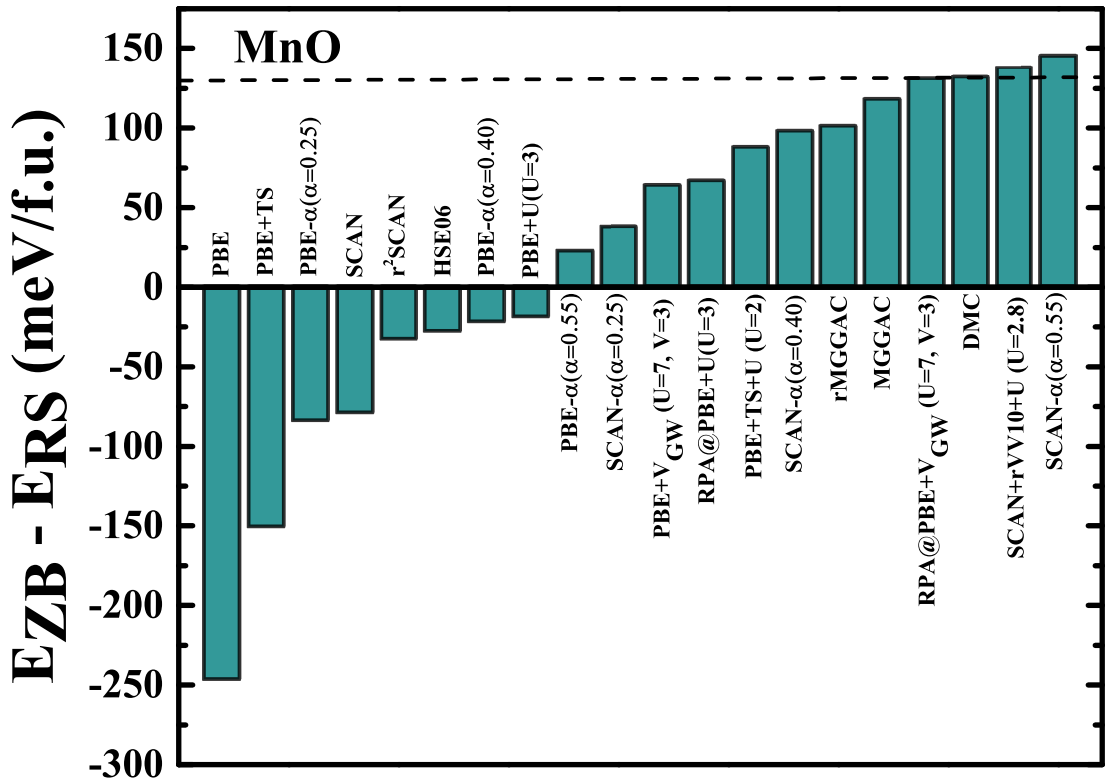}
\caption{Total energy difference $\Delta E=E_{zb}^{\text{AF1}}-E_{rs}^{\text{AF2}}$ (in meV/f.u.) between the $zb$-AF1 and $rs$-AF2 phases of MnO. A positive energy difference indicates that $rs$-AF2 is energetically more stable than $zb$-AF1 (as experimentally determined). The MGGAC, rMGGAC, r$^2$SCAN, and onsite hybrids results are calculated for this work. The results obtained with the other methods are from Refs.~\onlinecite{schiller2015phase,peng2013polymorphic,peng2017synergy}.}
\label{phase1}
\end{figure*}

\begin{figure*}
\includegraphics[width=17 cm, height=13 cm]{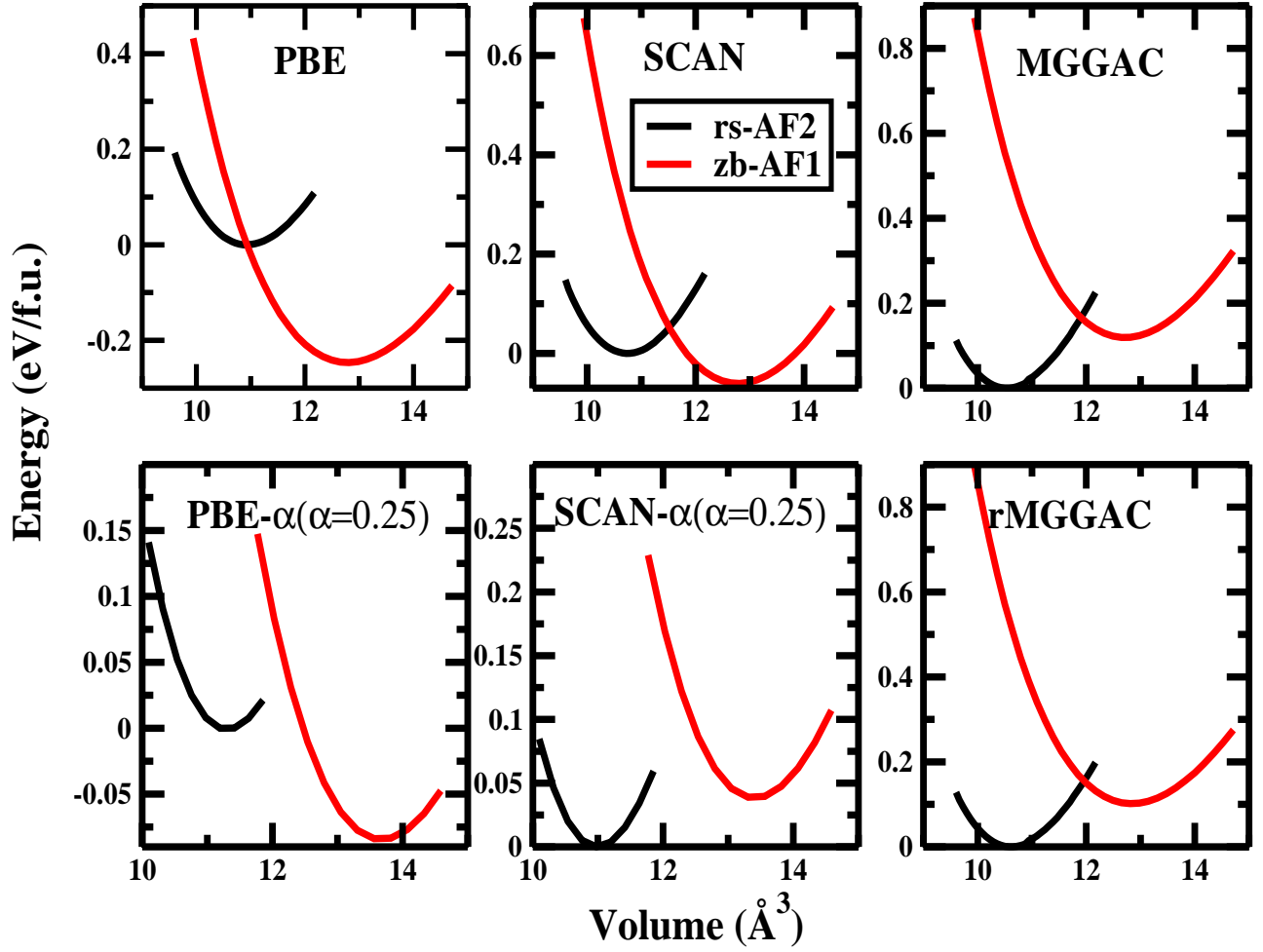}
\caption{Total energy versus volume (in \AA$^{3}$ per atom) for the $rs$-AF2 and $zb$-AF1 phases of MnO obtained from different methods. The total energy is set to zero at the minimum of the curve of the $rs$-AF2 phase.}
\label{ev-plot}
\end{figure*}

\begin{figure}
\includegraphics[width=10 cm, height=7 cm]{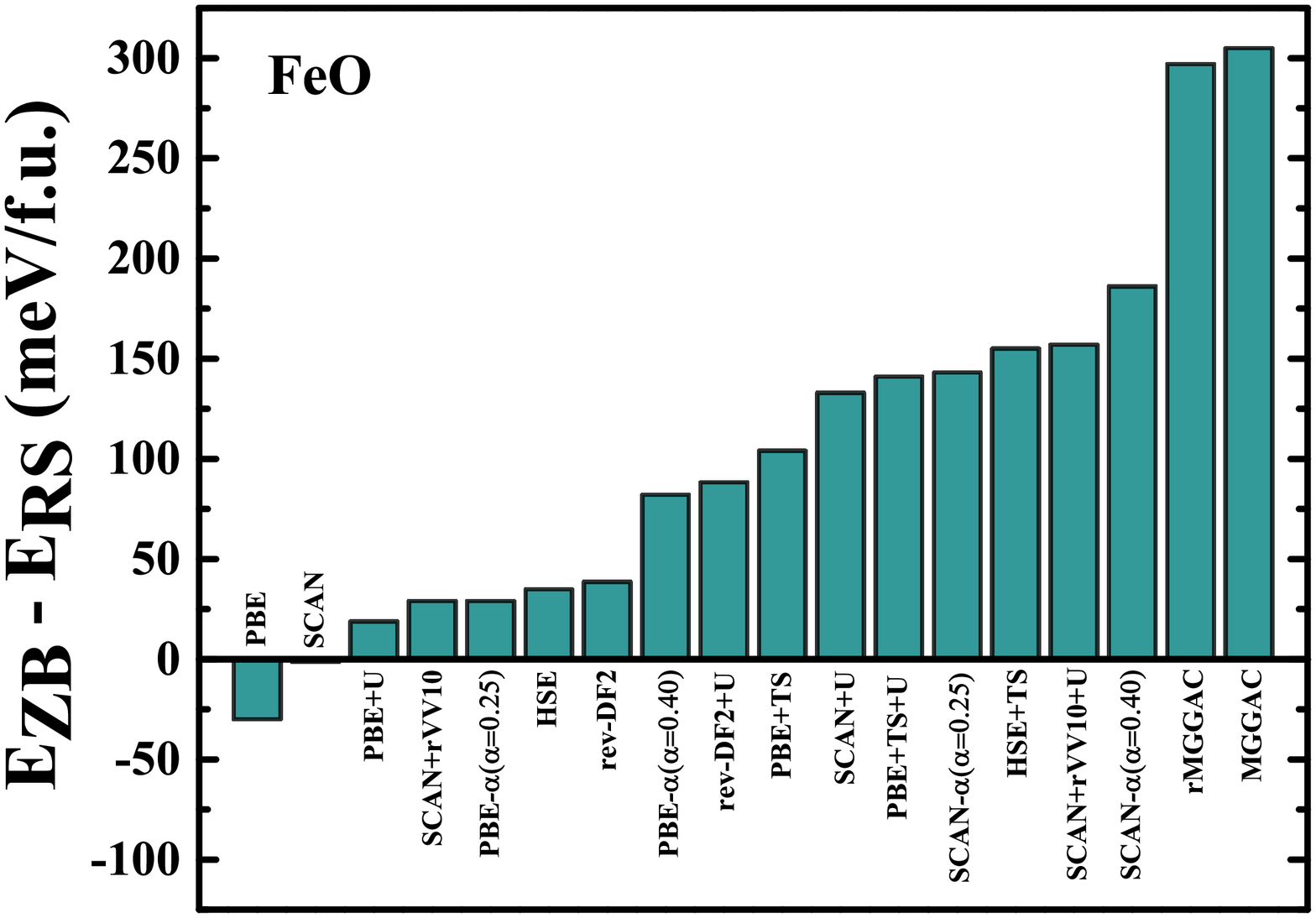}
\includegraphics[width=10 cm, height=7 cm]{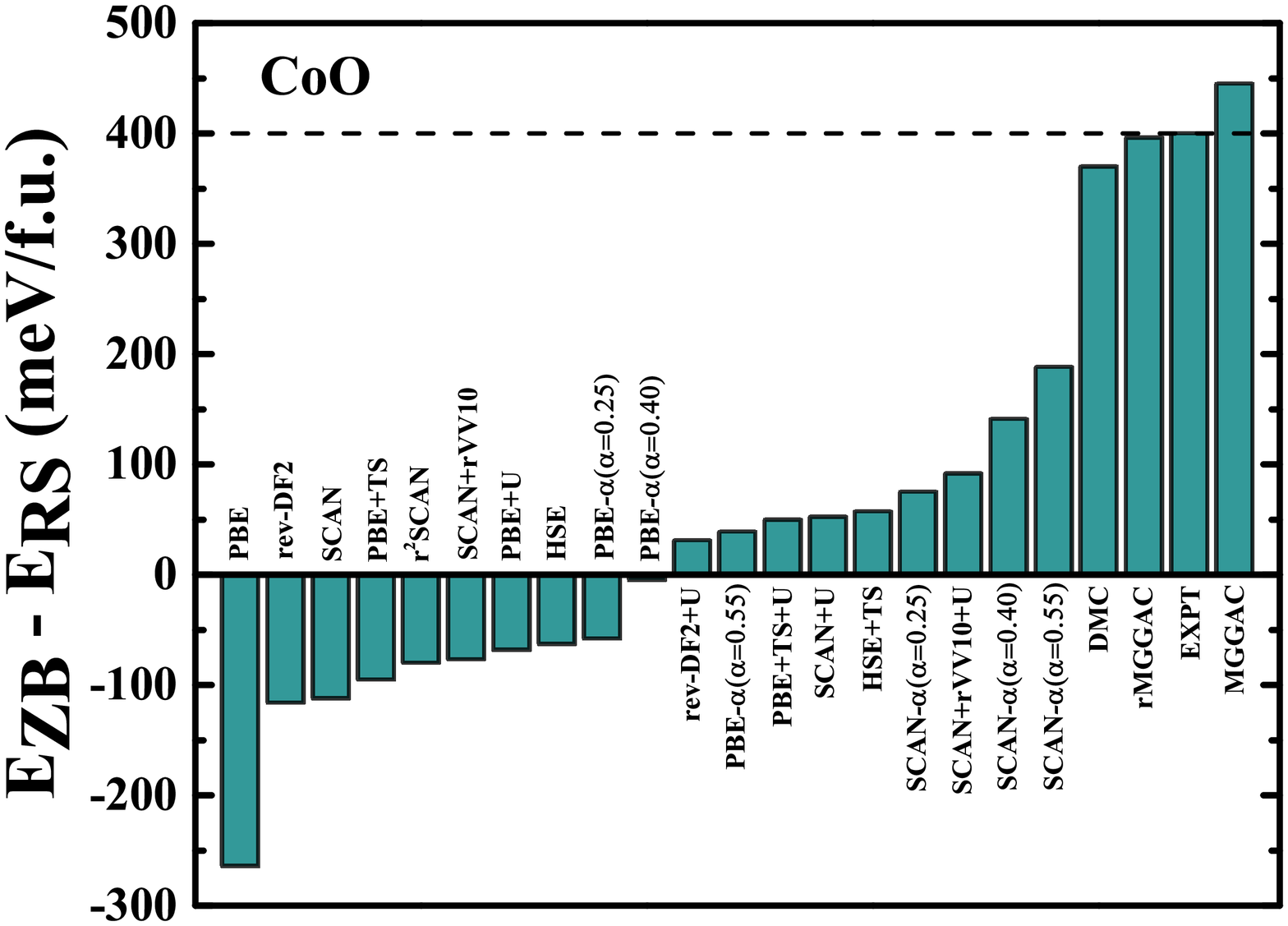}
\includegraphics[width=10 cm, height=7 cm]{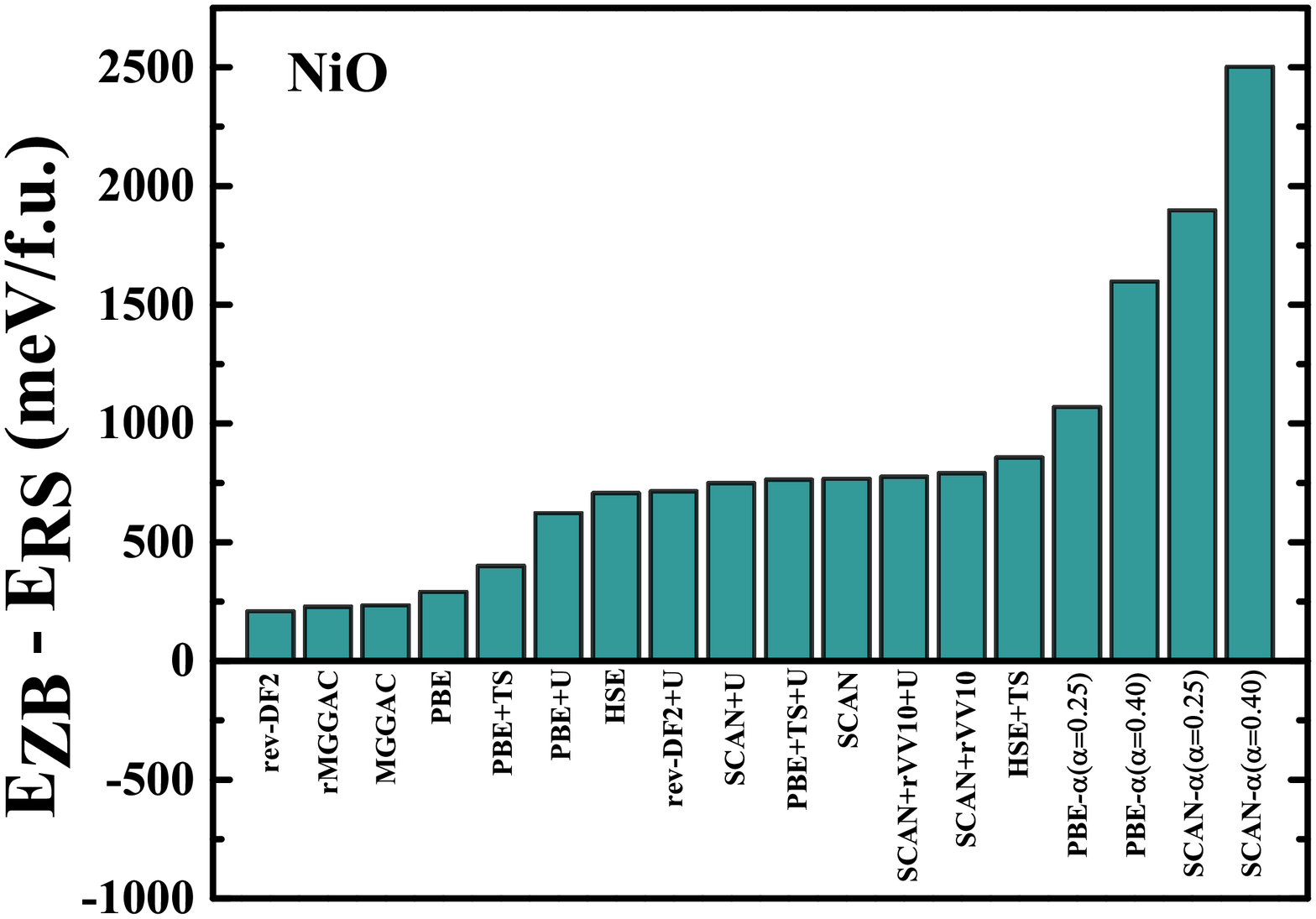}
\caption{Total energy difference $\Delta E=E_{zb}^{\text{AF1}}-E_{rs}^{\text{AF2}}$ (in meV/f.u.) between the $zb$-AF1 and $rs$-AF2 phases of FeO (upper panel), CoO (middle panel), and NiO (lower panel). A positive energy difference indicates that $rs$-AF2 is energetically more stable than $zb$-AF1. The MGGAC, rMGGAC, r$^2$SCAN, and onsite hybrids results are calculated for this work. See Refs.~\onlinecite{schiller2015phase,peng2013polymorphic,peng2017synergy} for the detailed results obtained with the other methods.}
\label{phase2}
\end{figure}

The calculated total energy difference $\Delta E=E_{zb}^{\text{AF1}}-E_{rs}^{\text{AF2}}$ and stability of the $zb$-AF1 phase relative to the experimentally determined ground-state $rs$-AF2 phase of MnO are shown in Table~\ref{tab-rel1} and Fig.~\ref{phase1}. The results obtained with PBE, (r$^{2}$)SCAN, (r)MGGAC, and the onsite hybrids are obtained for the present work and are compared to results from the literature obtained with other methods. As mentioned in Sec.~\ref{introduction}, methods from various levels of theory, including semilocal, vdW-corrected, and high-level correlation methods, have been used and proposed in previous works to correctly describe the relative phase stability of MnO and to provide estimates of $\Delta E$. In the present work, we consider the DMC value $\Delta E^{\text{DMC}}=132$~meV/f.u. from Ref.~\onlinecite{schiller2015phase} as the reference benchmark.

We mention again that RPA correctly leads to $rs$-AF2 as the ground-state phase \cite{peng2013polymorphic}. Compared to DMC, the value $\Delta E=67$~meV/f.u. from RPA@PBE+$U$ ($U=3$~eV) (PBE+$U$ orbitals used as input to RPA) is too small by a factor of 2, while $\Delta E=131$~meV/f.u from RPA@PBE+$V_{GW}$ ($U=7$~eV and $V=3$~eV) matches perfectly the DMC value (in PBE+$V_{GW}$ a nonlocal external potential $V_{GW}$ on Mn $d$ orbitals is used on top of the onsite $U$). In Ref.~\onlinecite{peng2017synergy} it is shown that the correct phase ordering can also be obtained by adding vdW (Tkatchenko-Scheffler (TS) \cite{TkatchenkoPRL09} or rVV10 \cite{doi:10.1063/1.3521275}) and Hubbard $U$ (obtained from linear-response approach) corrections to a semilocal functional. The proposed PBE+TS+$U$ ($U = 3.2$~eV)~\cite{peng2017synergy} and SCAN+$r$VV10+$U$ ($U = 2.8$~eV)~\cite{peng2017synergy} lead to $\Delta E=88$ and 138~meV/f.u., respectively, the latter value agreeing very well with DMC.

However, as clearly visible in Fig.~\ref{phase1}, there are also a certain number of other popular methods that fail in predicting the correct ground-state of MnO. Besides plain PBE, it is also the case with PBE+$U$ with a small value of $U$ (3~eV), the hybrid functional HSE06, and PBE+TS \cite{peng2013polymorphic}. The meta-GGAs (r$^{2}$)SCAN also lead to the wrong phase for the ground state, however the two other meta-GGAs tested in this work, MGGAC and rMGGAC, lead to positive energy differences, indicating that $rs$-AF2 is more stable than $zb$-AF1. Thus, MGGAC and rMGGAC are successful in predicting correctly the polymorphic energy ordering of the MnO phases. These are very interesting results, in particular when considering that no Hubbard $U$ or vdW corrections are added to (r)MGGAC. Actually, as evident from the results, MGGAC and rMGGAC give the best performance among the semilocal methods. They lead to values for $\Delta E$ of $118$ and $101$ meV/f.u., respectively,
which agree quite well with the reference value 132~meV/f.u. from DMC.

Regarding the performance of the onsite hybrids PBE-$\alpha$ and SCAN-$\alpha$, also tested for the present work, we can see that the energetic ordering of the $rs$-AF2 and $zb$-AF1 phases depends on the amount of the Hartree-Fock exchange $\alpha$. For a correct energetic ordering, PBE-$\alpha$ requires a value of $\alpha$ larger than 0.40, which is quite high and clearly larger than the standard value $\alpha=0.25$. For SCAN-$\alpha$, $\alpha=0.25$ is already sufficient. The observed trend with the onsite hybrids is that the more $\alpha$ is large, the more the energetic ordering will go in the right direction. This is in agreement with the results from Schiller \textit{et al}. \cite{schiller2015phase} who considered hybrid functionals based on PBE (however, they found that $rs$-AF2 becomes more stable already at $\alpha=0.10$). This behavior is also similar to what is obtained with DFT+$U$, with $U$ playing the same role as $\alpha$~\cite{kanan2012band,peng2013polymorphic}.

In Fig.~\ref{ev-plot}, we show the total energy versus volume for the two phases of MnO obtained with various methods. As discussed above, unlike for the PBE and SCAN methods the total-energy volume curve of the $rs$-AF2 phase is below that for the $zb$-AF1 phase when MGGAC and rMGGAC are used.

\begin{table*}
\caption{Total energy difference $\Delta E=E_{zb}^{\text{AF1}}-E_{rs}^{\text{AF2}}$ (in meV/f.u.) between the $zb$-AF1 and $rs$-AF2 phases of MnO obtained non-self-consistently with different functionals using the PBE orbitals and electron density. The calculations are performed using the all-electron code WIEN2k.}
\begin{ruledtabular}
\begin{tabular}{lccccc}
&PBE&SCAN@PBE&r$^2$SCAN@PBE&MGGAC@PBE&rMGGAC@PBE\\
\hline
$\Delta E$&-243&-87&-57&109&84\\
\end{tabular}
\end{ruledtabular}
\label{tab-non-scf}
\end{table*}

\begin{table*}
\caption{Lattice constant $a_0$, bulk modulus $B_0$, band gap $E_g$, and magnetic moment $\mu$ of the Mn atom of the $rs$-AF2 phase of MnO obtained from different methods. The calculations with the semilocal functionals are performed using the VASP code, while the onsite hybrid calculations are done using the WIEN2k code.}
\begin{ruledtabular}
\begin{tabular}{lccccc}
Method & $a_0$ (\AA) & $B_0$ (GPa) & $E_g$ (eV) & $\mu$ ($\mu_{\text{B}}$) \\
\hline
PBE                           & 4.438 & 149 & 0.72 & 4.38 \\
SCAN                          & 4.411 & 163 & 1.47 & 4.49 \\
r$^{2}$SCAN                   & 4.418 & 166 & 1.52 & 4.50 \\
MGGAC                         & 4.381 & 185 & 1.77 & 4.52 \\
rMGGAC                        & 4.392 & 178 & 1.70 & 4.52 \\
PBE-$\alpha$ ($\alpha=0.25$)  & 4.488	&144		&1.21	&4.65 \\
SCAN-$\alpha$ ($\alpha=0.25$)\footnotemark[1] & 4.447	&161&	N/A		& N/A    \\
PBE-$\alpha$ ($\alpha=0.40$)  & 4.512	&141		&1.34	&4.74 \\
SCAN-$\alpha$ ($\alpha=0.40$)\footnotemark[1] & 4.465	&158		&   N/A 	&   N/A  \\
PBE-$\alpha$ ($\alpha=0.55$)  & 4.531	&139		&1.45	&4.80 \\
SCAN-$\alpha$ ($\alpha=0.55$)\footnotemark[1] & 4.482	&156		&  N/A  	& N/A     \\
Expt. \footnotemark[2] & 4.4365, 4.4315, 4.4302 & 149.6, 146.7, 148/144 & 3.6-3.9 & 4.58 \\
\end{tabular}
\footnotetext[1] {Since the SCAN functional in SCAN-$\alpha$ was applied non-self-consistently (the PBE potential was used) the values of $E_g$ and $\mu$ are omitted.}\\
\footnotetext[2]{See Ref.~\onlinecite{schron2010energetic} and references therein.}
\end{ruledtabular}
\label{tab-gs}
\end{table*}

\begin{table*}
\caption{Values of $E_{\text{xc}}$ in the $rs$-AF2 and $zb$-AF1 phases of MnO calculated with the SCAN and MGGAC functionals. The results are obtained non-self-consistently (with the PBE electron density) at the SCAN equilibrium geometry using the WIEN2k code. The total value in the cell is decomposed into the Mn and O atomic spheres and interstitial. The differences between the two phases are also shown, as well as their difference between the two functionals (last column). The units are in Ry/f.u. for columns 1, 2, 4, and 5 and in meV/f.u. for columns 3, 6, and 7.}
\begin{ruledtabular}
\begin{tabular}{lccccccc}
\multicolumn{1}{l}{} & \multicolumn{3}{c}{SCAN} &
\multicolumn{3}{c}{MGGAC} &
\multicolumn{1}{c}{SCAN$-$MGGAC} \\
\cline{2-4}\cline{5-7}
 & $rs$ & $zb$ & $rs-zb$ & $rs$ & $zb$ & $rs-zb$ & $rs-zb$ \\
 \hline
Cell         &-122.693& -122.595& -1324& -123.077& -122.966& -1515& 191\\
Mn           &-104.304& -104.246& -786 & -104.470& -104.409& -829 &  43\\
O            &-16.887 & -16.934 &  646 & -17.105 & -17.152 &  649 &  -3\\
Interstitial &-1.502  & -1.415	& -1183& -1.502  & -1.404  & -1335& 152\\
\end{tabular}
\end{ruledtabular}
\label{table_decomp}
\end{table*}

\begin{figure}[!ht]
\includegraphics[width=\columnwidth]{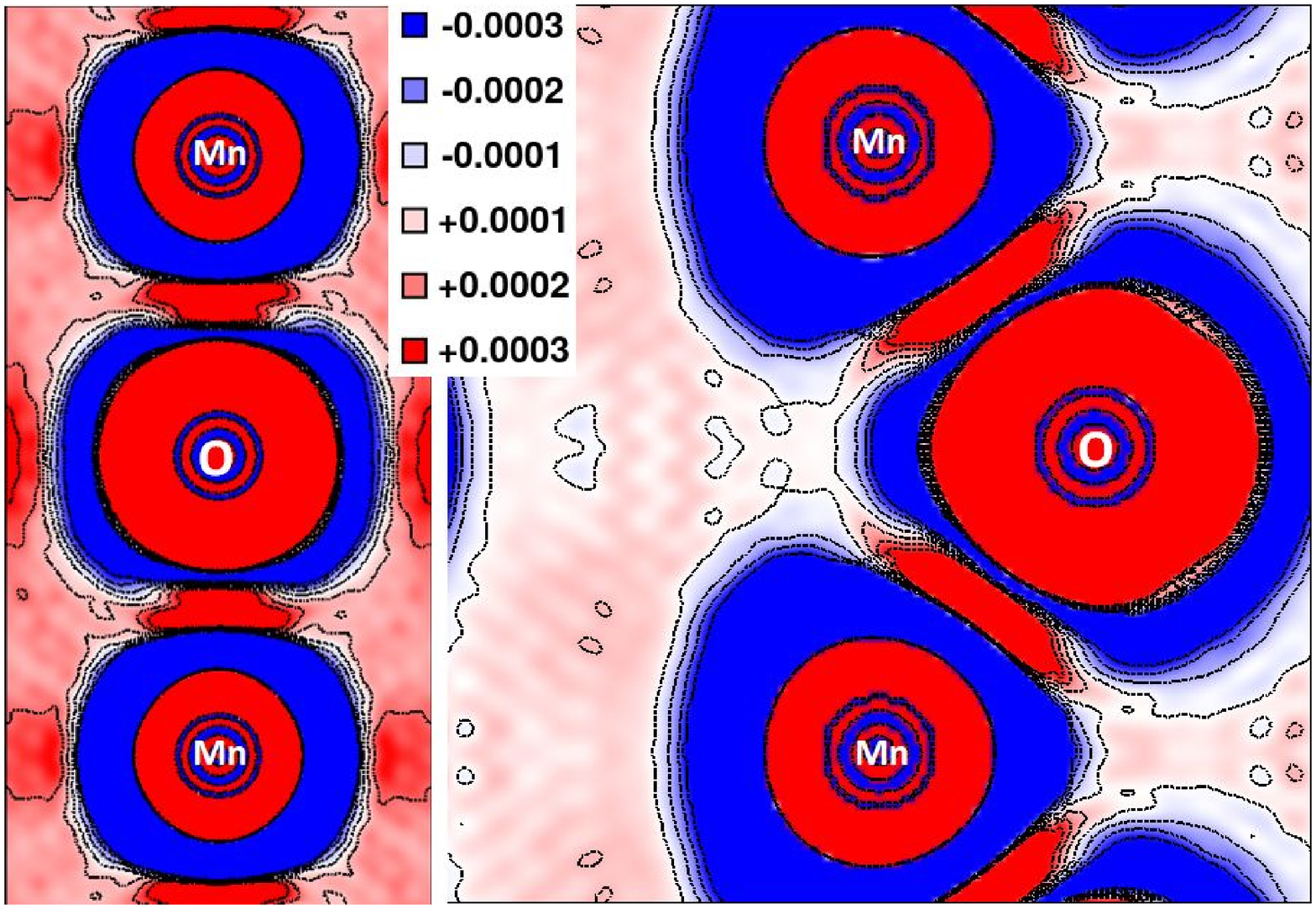}
\caption{\label{fig_exc}Two-dimensional plot in MnO of the difference $\epsilon_{\text{xc}}^{\text{SCAN}}-\epsilon_{\text{xc}}^{\text{MGGAC}}$ between the SCAN and MGGAC xc energy density [see Eq.~(\ref{Exc})] obtained with WIEN2k from non-self-consistent calculations using the PBE electron density. The left and right panels show the (110) plane of $rs$-AF2 and the (001) plane of $zb$-AF1, respectively. The blue and red regions correspond to negative and positive values, respectively.}
\end{figure}

\begin{figure}[!ht]
\includegraphics[width=\columnwidth]{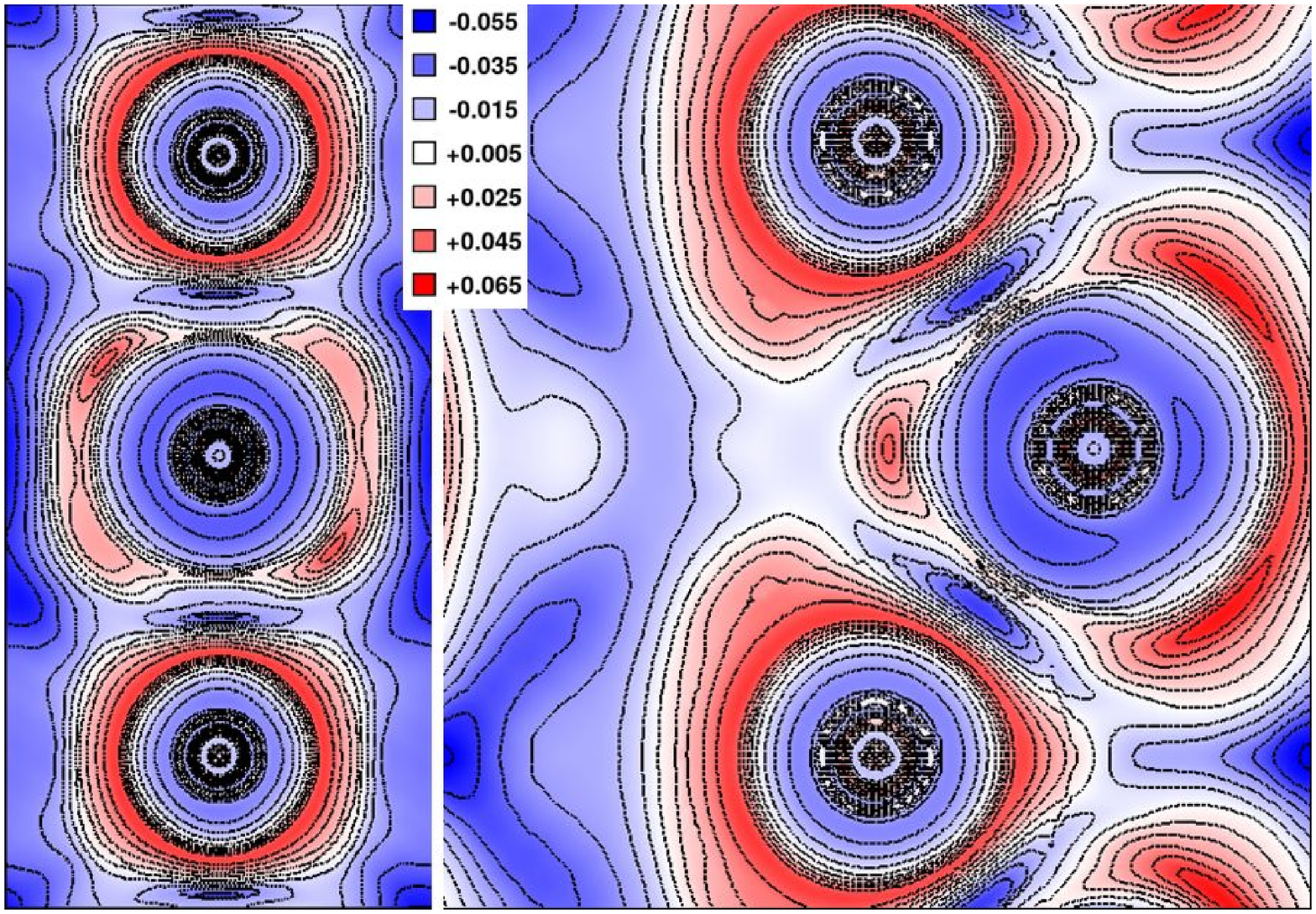}
\caption{\label{fig2D_Fxc}Two-dimensional plot in MnO of the difference $F_{\text{xc}}^{\text{SCAN}}-F_{\text{xc}}^{\text{MGGAC}}$ between the SCAN and MGGAC xc enhancement factors [see Eq.~(\ref{Exc})] obtained with WIEN2k from non-self-consistent calculations using the PBE electron density. The left and right panels show the (110) plane of $rs$-AF2 and the (001) plane of $zb$-AF1, respectively. The blue and red regions correspond to negative and positive values, respectively. The atoms are the same as indicated in Fig.~\ref{fig_exc}.}
\end{figure}

\begin{figure}[!ht]
\includegraphics[width=\columnwidth]{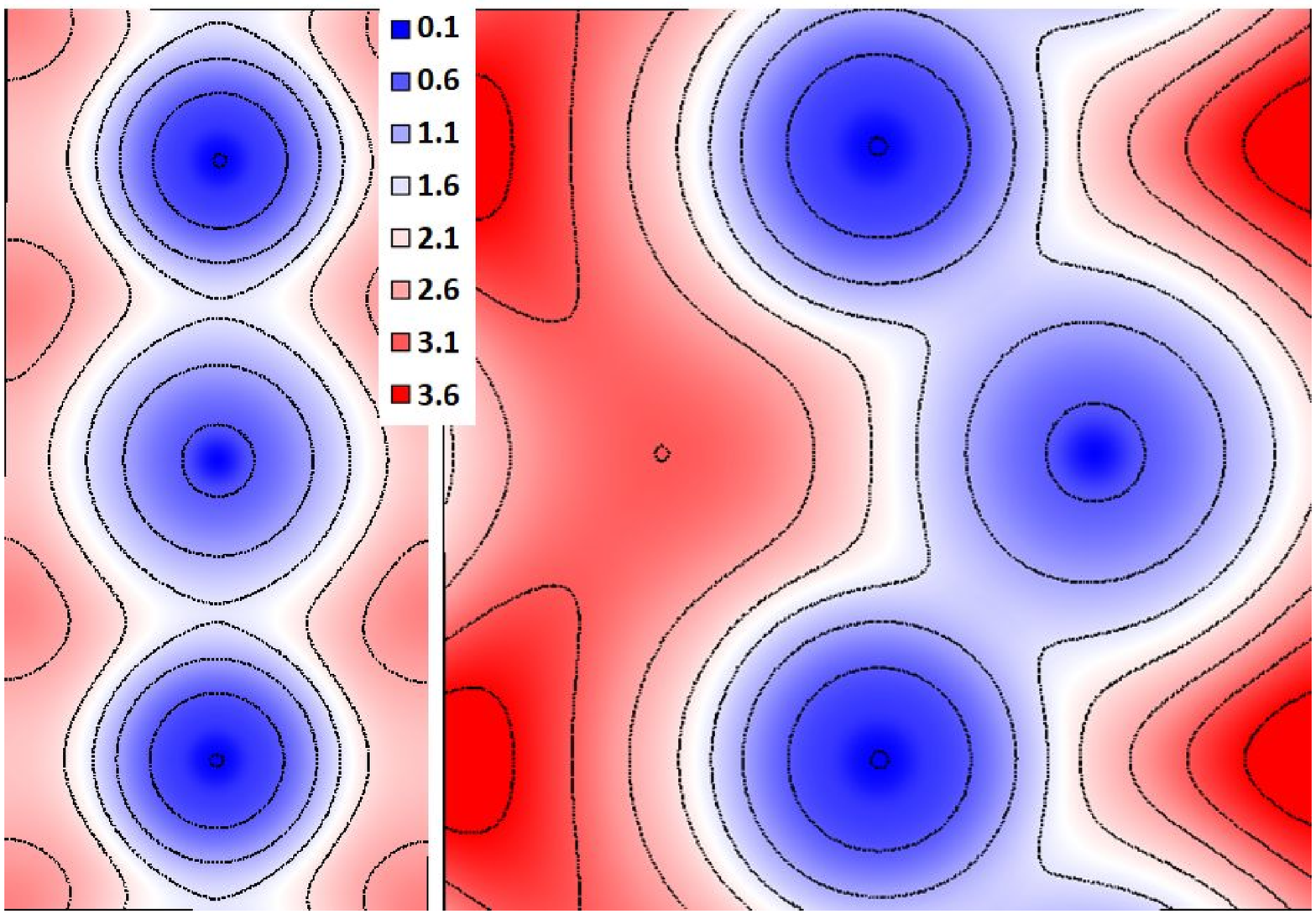}
\includegraphics[width=\columnwidth]{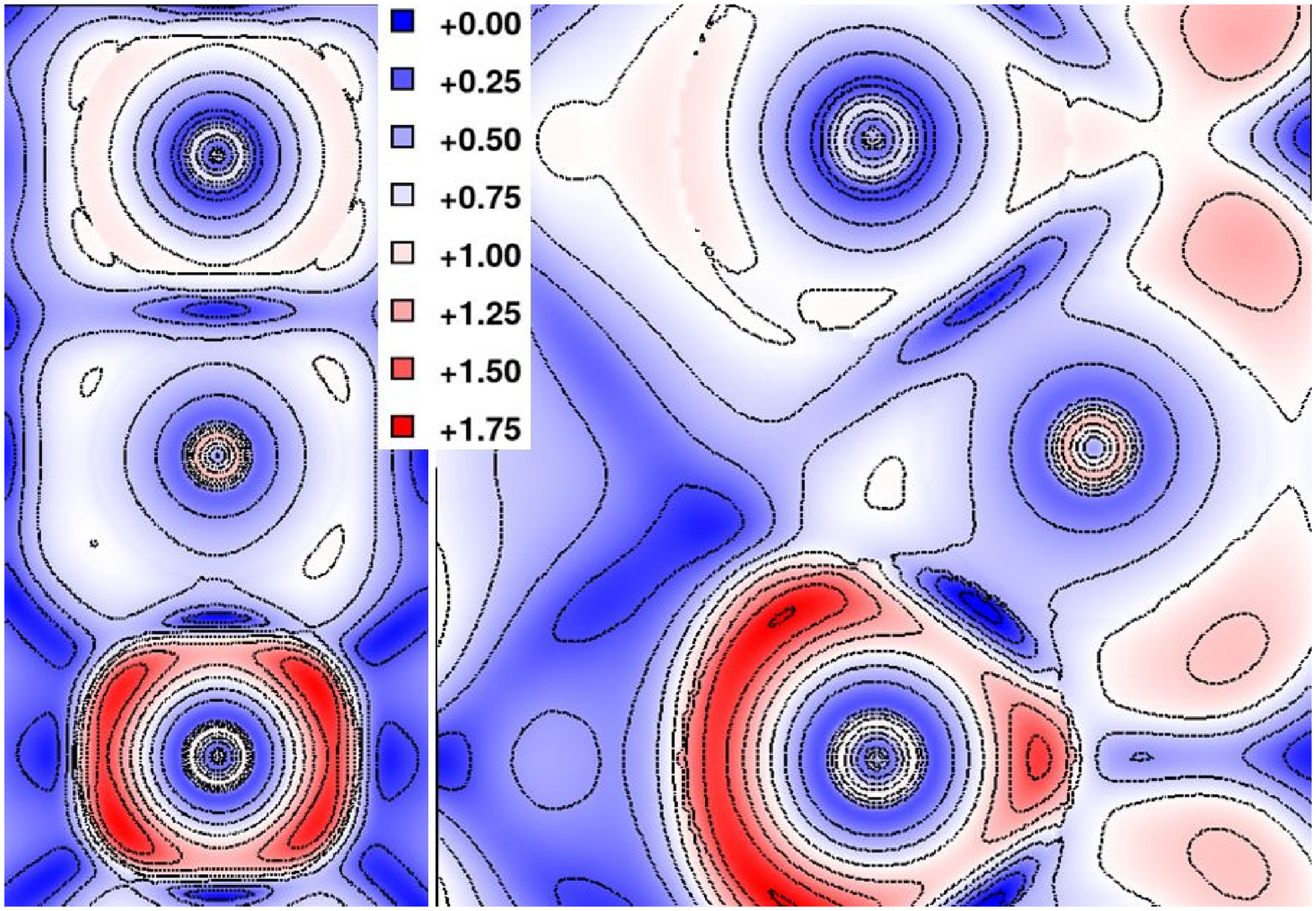}
\includegraphics[width=\columnwidth]{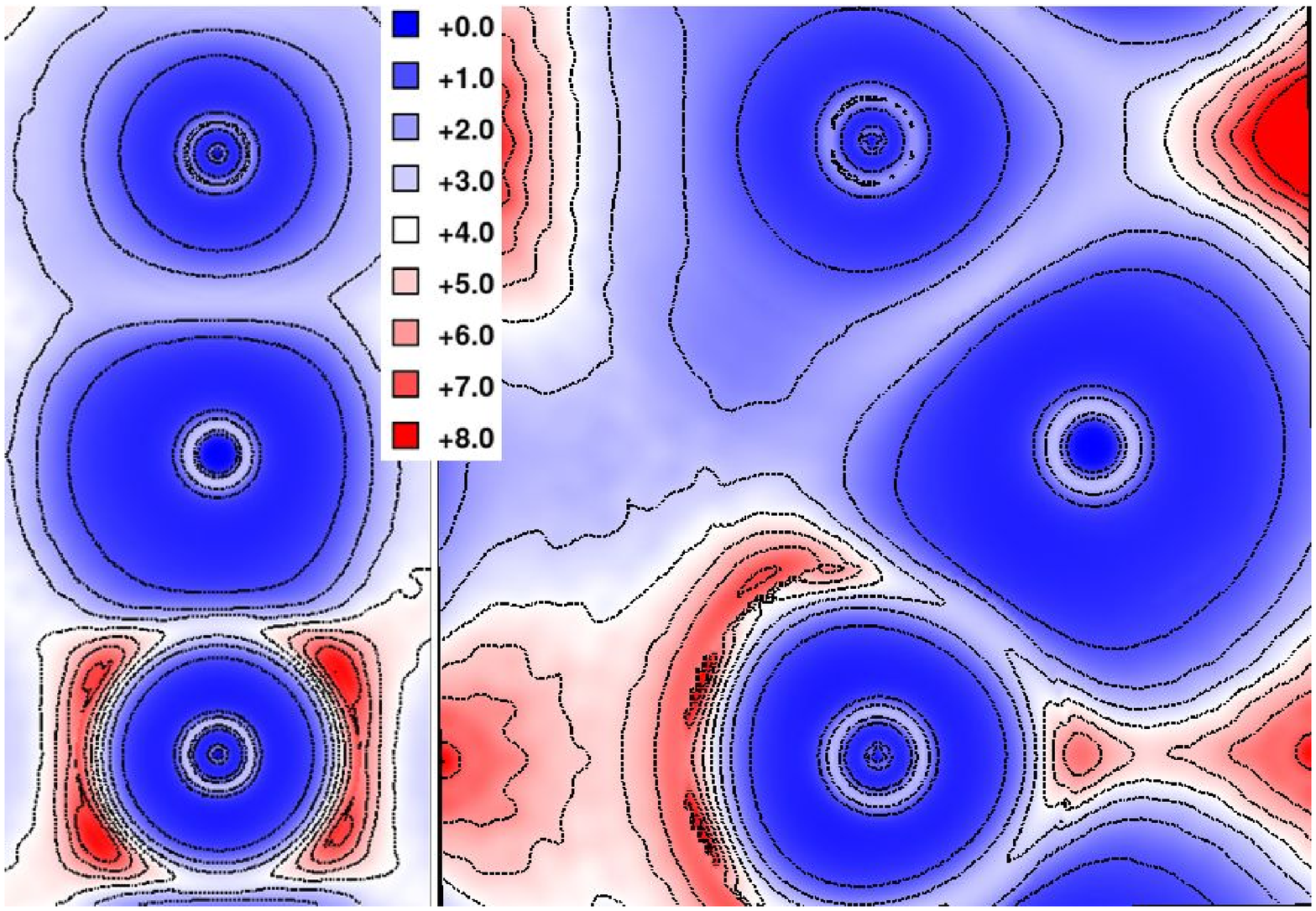}
\caption{\label{fig_MnO_plots}Two-dimensional plots in MnO of $r_s$ (upper panels), the $\sigma$-spin reduced density gradient $s_{\sigma}$ (middle panels), and the $\sigma$-spin iso-orbital indicator $\alpha_{\sigma}^{\text{iso}}$ (lower panels). The left and right panels show the (110) plane of $rs$-AF2 and the (001) plane of $zb$-AF1, respectively. The atoms are the same as indicated in Fig.~\ref{fig_exc}.}
\end{figure}

The (r$^{2}$)SCAN and (r)MGGAC results discussed so far are obtained self-consistently using the Vienna Ab initio Simulation Package (VASP) code~\cite{PhysRevB.47.558,PhysRevB.54.11169,KRESSE199615,PhysRevB.59.1758}. It may be of interest to determine how important is self-consistency for the results, in particular in the case of (r)MGGAC, which lead to very good results. It is generally believed that in most cases self-consistency plays only a minor role and that using for instance the PBE electron density and orbitals (instead of those from (r)MGGAC) would not affect much the total energy computed with the (r)MGGAC energy functional. However, it may not be always the case, as discussed in Ref.~\onlinecite{PhysRevLett.111.073003} for the closely related density-driven error. To make this point more clear for the results in the present work, we also calculated $\Delta E$ for (r)MGGAC and (r$^{2}$)SCAN using the PBE orbitals and electron density. These non-self consistent calculations are performed using the all-electron code WIEN2k\cite{WIEN2k,BlahaJCP20} and the results are reported in Table~\ref{tab-non-scf}. We can see that by using the PBE density and orbitals the trends do not change, i.e., the ordering of the phases with (r)MGGAC is still correct, whereas it is not the case with (r$^2$)SCAN. There are some differences between the VASP self-consistent and WIEN2k non-self-consistent results, however they are unimportant for the conclusion. Overall, this indicates that the correct behavior of (r)MGGAC is essentially functional-driven.

Next, we consider in detail the results for the structural, electronic, and magnetic properties of $rs$-AF2 MnO, calculated with the semilocal and onsite hybrid functionals. Our results are summarized in Table~\ref{tab-gs} and are compared to experimental results (see Ref.~\onlinecite{schron2010energetic} and references therein). The experimental equilibrium lattice constant $a_{0}\sim4.43$~\AA~is best reproduced by PBE. (r$^{2}$)SCAN and SCAN-$\alpha$ ($\alpha=0.25$) are also pretty accurate. For the bulk modulus $B_{0}$, PBE is again the best method, and PBE-$\alpha$ can be also very accurate depending on the value of $\alpha$, while (r)MGGAC give values of $B_{0}$ that are too large by about 30~GPa.

The experimental band gap of 3.6$-$3.9~eV is strongly underestimated by at least 2~eV by all methods in Table~\ref{tab-gs}. In a recent study \cite{PhysRevMaterials.2.023802} it was shown that among the fast semilocal DFT methods only the GLLB-SC \cite{KuismaPRB10} and Sloc \cite{FinzelIJQC17} functionals are able to give band gaps of MnO similar to experiment. Other studies \cite{rodl2009quasiparticle,schiller2015phase} have shown that $GW$ can be accurate depending on the input orbitals, while DMC gives a band gap that is too large by nearly 1~eV. The magnetic moment $\mu$ on the Mn atom was calculated inside the atomic basin as defined by the quantum theory of atoms in molecules of Bader\cite{Bader90,BaderCR91}. The experimental value of 4.58~$\mu_{B}$ is best reproduced by the MGGAC, rMGGAC, and PBE-$\alpha$($\alpha=0.25$) functionals that give 4.52, 4.52, and 4.65~$\mu_{\text{B}}$, respectively.

Finally, we show in Fig.~\ref{phase2} the results for the other antiferromagnetic TMOs considered in this work: FeO, CoO, and NiO. As for MnO, their ground-state phase is $rs$-AF2 \cite{RothPR58}, and we can see that this is correctly predicted by the MGGAC and rMGGAC functionals.

Neither experimental data nor values obtained from highly accurate methods like DMC or RPA seem to be available for FeO. Therefore, a comparison of the values of $\Delta E$ can be made only between DFT methods. As in the cases of MnO and CoO (see below), (r)MGGAC lead to the largest positive values of $\Delta E$, too. They are clearly larger than for all other methods. Next come the onsite hybrid SCAN-$\alpha$ and $U$- and/or vdW-corrected functionals like SCAN+rVV10+$U$ or HSE+TS. Only PBE and SCAN lead erroneously to $zb$-AF1 as the ground-state phase, while a small amount of Hartree-Fock exchange $\alpha$ or small value of $U$ is enough to get the correct ground-state phase $rs$-AF2.

For CoO, the experimental and DMC values (see Ref.~\onlinecite{saritas2018relative} and references therein) are around $400$~meV/f.u. and are reproduced very accurately by (r)MGGAC, as visible on Fig.~\ref{phase2}. Considering the onsite hybrid functionals based on SCAN, using larger values of $\alpha$ leads to better agreement with experiment and DMC, however the values are twice too small even with $\alpha=0.55$. Values from the literature \cite{schiller2015phase,peng2013polymorphic,peng2017synergy,saritas2018relative} obtained with other methods are also shown in Fig.~\ref{phase2}. As for MnO, the popular methods PBE, SCAN, and HSE06 fail since they predict $zb$-AF1 to be more stable than $rs$-AF2, but adding a Hubbard $U$ and/or vdW correction helps to get the correct trend.

As for FeO, no reference data is available for NiO. By inspecting the DFT results, we can see that the situation is quite different compared to MnO, FeO, and CoO. Firstly, all functionals, without exception, lead to $rs$-AF2 as the ground-state phase. Secondly, the MGGAC and rMGGAC values are basically the smallest in magnitude, while the reverse was obtained for MnO, FeO, and CoO. The largest values of $\Delta E$ are obtained with the onsite hybrids SCAN-$\alpha$ and PBE-$\alpha$, and are much larger than what is obtained with the other methods. It should also be noted that the calculated energetic differences between the $rs$-AF2 and $zb$-AF1 phases are much larger than for the other studied TMOS by nearly one order of magnitude.

In order to show that the (r)MGGAC functionals lead to the correct energy ordering also for other phases of the studied systems, we considered the $wz$-AF1 phase \cite{schron2010energetic}. The values of $\Delta E^{wz-rs}=E_{wz}^{\text{AF1}}-E_{rs}^{\text{AF2}}$ are presented in Tables~S2 and S3 of Ref.~\onlinecite{support}. For the four systems the (r)MGGAC values of $\Delta E^{wz-rs}$ are positive, which is the correct trend. Reference (experimental and DMC) values for $\Delta E^{wz-rs}$ are available only for CoO \cite{saritas2018relative}, and we can see that the magnitude of $\Delta E^{wz-rs}$ obtained with (r)MGGAC is too large by roughly $\sim200$~meV/f.u. compared to the reference values.

At this point one may wonder why the (r)MGGAC functionals work much better than PBE and (r$^{2}$)SCAN for the relative stability of the phases of the studied TMOs. In order to address this question, we performed calculations with the WIEN2k code and decomposed $E_{\text{xc}}$ into contributions from the atomic spheres and the interstitial. We show in Table~\ref{table_decomp} the total as well as the decomposed values of $E_{\text{xc}}$ obtained from SCAN and MGGAC for the $rs$-AF2 and $zb$-AF1 phases of MnO. MGGAC stabilizes $rs$-AF2 over $zb$-AF1 by 1515~meV/f.u., while SCAN gives a smaller value, 1324~meV/f.u., and this difference of 191~meV/f.u. leads as discussed above to the correct ground state with MGGAC but not with SCAN. By looking from which region this difference comes, there is virtually no contribution from the O atom, a very small one from Mn, while most of the stabilization comes from the interstitial region. Of course the exact values depend weakly on the chosen atomic sphere sizes (integration ranges). Therefore, we show in Fig.~\ref{fig_exc} the difference $\epsilon_{\text{xc}}^{\text{SCAN}}-\epsilon_{\text{xc}}^{\text{MGGAC}}$ between the SCAN and MGGAC xc energy density [the integrand in Eq.~(\ref{Exc})]. There are large positive and negative differences inside the atomic sphere since the corresponding density is large, but as shown in Table~\ref{table_decomp} these differences are very similar for the two phases and cancel to a large extent. However, in the interstitial regions we can clearly see that this difference is close to zero in $zb$-AF1, but much more positive in $rs$-AF2. A positive difference $\epsilon_{\text{xc}}^{\text{SCAN}}-\epsilon_{\text{xc}}^{\text{MGGAC}}$ means a more negative integrand of $E_{\text{xc}}$ for MGGAC and thus a stabilization of the $rs$-AF2 structure. An examination of the difference of the enhancement factors $F_{\text{xc}}^{\text{SCAN}}-F_{\text{xc}}^{\text{MGGAC}}$ (Fig.~\ref{fig2D_Fxc}) shows partially a similar picture (with opposite sign) as in Fig.~\ref{fig_exc}, namely large positive and negative differences inside the atomic spheres, which at the end are similar for both $rs$-AF2 and $zb$-AF1 phases and thus cancel. On the other hand this difference is more negative in the interstitial of the $rs$-AF2 phase and together with a larger density gives the dominant contribution.

Since now we know from which region (the interstitial) the effect comes, we can mention what are the values of $r_{s}$, and the $\sigma$-spin $s_{\sigma}$ and $\alpha_{\sigma}^{\text{iso}}$ in this region (see Fig.~\ref{fig_MnO_plots}). The value of $r_{s}$ corresponding to the red interstitial region of the $rs$-AF2 phase in $\epsilon_{\text{xc}}^{\text{SCAN}}-\epsilon_{\text{xc}}^{\text{MGGAC}}$ in Fig.~\ref{fig_exc}, or equivalently in blue in Fig.~\ref{fig2D_Fxc}, is around 2~Bohr, while in the large interstitial region of the $zb$-AF1 phase it is more around 3~Bohr, which corresponds to lower density than in $rs$-AF2. $s_{\sigma}$ is quite small (from 0 to 0.5), while $\alpha_{\sigma}^{\text{iso}}$ is moderately large (from 2 to 3). However, it should be also noted that $\alpha_{\sigma}^{\text{iso}}$ can reach values up to 8 in some regions just outside of the $3d$ density (where also $s_{\sigma}$ increases up to 1.7) and in the interstitial of $zb$-AF1 (where $s_{\sigma}$ is very small).

We can now compare the corresponding values of the enhancement factors $F_{\text{xc}}$ of SCAN and MGGAC shown in Fig.~\ref{fig_Fxc}. For the relevant values in the interstitial region of $r_{s}$ (in the range 2-3), $s$ (below $\sim0.7$), and $\alpha_{\sigma}^{\text{iso}}$ (above $\sim2$), we can see that $F_{\text{xc}}^{\text{MGGAC}}$ is more positive than $F_{\text{xc}}^{\text{SCAN}}$. Thus, since $\epsilon_{\text{xc}}$ is proportional to $-n^{4/3}F_{\text{xc}}$ and the electron density $n$ in the interstitial region is larger in $rs$-AF2 than in $zb$-AF1, then $\epsilon_{\text{xc}}^{\text{MGGAC}}-\epsilon_{\text{xc}}^{\text{SCAN}}$ (and $E_{\text{xc}}^{\text{MGGAC}}-E_{\text{xc}}^{\text{SCAN}}$, see Table~\ref{table_decomp}) is more negative in $rs$-AF2 than in $zb$-AF1, leading to the observed stabilization of $rs$-AF2 with MGGAC.

\section{\label{conclusions}Conclusions}

In this work, the ability of various DFT methods to predict the correct ground-state phase of antiferromagnetic transition-metal monoxides has been studied. The case of MnO has been considered in more details. The conclusions are the following. The meta-GGAs MGGAC and rMGGAC provide the correct energy ordering of the $rs$-AF2 and $zb$-AF1 phases of MnO, while it is not the case with the other popular functionals PBE, SCAN, and HSE06. Furthermore, the relative energies of the two phases are in very nice agreement with the values obtained with the highly accurate DMC and RPA methods. It should be underlined that the very good (r)MGGAC results have been obtained without addition of vdW or Hubbard $U$ correction. With other semilocal popular methods like PBE or SCAN, it is necessary to add a $U$ or vdW correction to obtain the correct ordering of the two phases. Since also the (r)MGGAC relative energy is very accurate, this may indicate that (r)MGGAC is more accurate in the atomic regions (where, alternatively, $U$ can also be added to improve the description), but also in the interstitial region (where a vdW correction may be helpful). We also showed in the present work that a relatively high amount of Hartree-Fock exchange (which plays the same role as $U$) in onsite hybrids is necessary to obtain the correct ordering of the phases of MnO.

Our results suggest that the semilocal (r)MGGAC functionals may be good alternatives to the costly DMC and RPA methods to predict the ground-state phase of strongly correlated systems. The low computational cost of semilocal methods is certainly helpful for studying potentially interesting functionalities of complex materials, such as heterostructure, cathode materials, or alloys, where various structural phases may compete.

\subsection{Computational details}

The calculations on MnO, FeO, CoO, and NiO with the r$^{2}$SCAN and (r)MGGAC meta-GGA functionals are performed using the Vienna Ab initio Simulation Package (VASP) code~\cite{PhysRevB.47.558,PhysRevB.54.11169,KRESSE199615,PhysRevB.59.1758}, which is based on the projector augmented wave method \cite{BlochlPRB94b,PhysRevB.59.1758}. The used pseudopotentials correspond to valence electron configurations $3d^64s^1$ for Mn, $3d^74s^1$ for Fe, $3d^84s^1$ for Co, $3d^{9}4s^1$ for Ni, and $2s^2p^4$ for O.

The calculations with the onsite hybrid functionals are done using the all-electron WIEN2k code \cite{WIEN2k,BlahaJCP20}, which is based on the augmented plane wave plus local orbitals method \cite{Singh,KarsaiCPC17}.

All VASP and WIEN2k calculations are done with parameters (size of basis set, \textbf{k}-mesh, etc.) that should lead to well converged results.

To calculate the equilibrium ground-state properties (total energy, volume, isothermal bulk modulus and its pressure derivative) we fitted the total energy versus volume data with the Birch-Murnaghan isothermal equation of state.

\begin{acknowledgements}
A.G. would like to thank INSPIRE fellowship, DST, India for financial support. The first-principle calculations within VASP are performed using the high performance computing (HPC) clusters of IITH, Hyderabad, KALINGA and NISERDFT high performance computing (HPC) clusters of NISER, Bhubaneswar.
\end{acknowledgements}


\twocolumngrid
\bibliography{reference.bib}
\bibliographystyle{apsrev4-1}

\end{document}